\documentclass{JHEP}

\usepackage{epsfig}
\usepackage{amsfonts}
\usepackage{graphicx,rotating}
\usepackage{psfrag}
\usepackage{bbm}
\usepackage{axodraw}
\usepackage{amssymb}

\newcommand{\bea}{\begin{eqnarray}}

\newcommand{\eea}{\end{eqnarray}}

\newcommand{\bean}{\begin{eqnarray*}}

\newcommand{\eean}{\end{eqnarray*}}

\newcommand{\non}{\nonumber}

\newcommand{\barray}{\begin{array}}
\newcommand{\eeqarray}{\end{eqnarray*}}
\newcommand{\beqarray}{\begin{eqnarray*}}
\newcommand{\earray}{\end{array}}
\newcommand{\bite}{\begin{itemize}}
\newcommand{\eite}{\end{itemize}}
\newcommand{\bmath}{\begin{displaymath}}
\newcommand{\emath}{\end{displaymath}}
\newcommand{\bdm}{\begin{displaymath}}
\newcommand{\edm}{\end{displaymath}}
\newcommand{\bd}{\begin{displaymath}}
\newcommand{\ed}{\end{displaymath}}

\newcommand{\be}{\begin{equation}}
\newcommand{\ee}{\end{equation}}

\newcommand{\beq}{\begin{equation}}
\newcommand{\eeq}{\end{equation}}
\newcommand{\bqa}{\begin{eqnarray}}
\newcommand{\eqa}{\end{eqnarray}}
\newcommand{\nl}{\nonumber \\}

\def\d#1{D_{#1}}
\def\tld#1{\tilde {#1}}

\def\Label#1{\label{#1}
 \smash{\hbox to0pt{\raise1ex\hbox{\tiny[#1]}\hss}}}

\title{NLO QCD corrections to tri-boson production}

\author{ T.~Binoth${}^a$, G.~Ossola${}^b$, C.~G.~Papadopoulos${}^b$, and R.~Pittau${}^{c}$ \\

\bigskip

${}^a$ The University of Edinburgh, School of Physics, Edinburgh EH9 3JZ, UK \\

\bigskip

${}^b$ Institute of Nuclear Physics, NCSR Demokritos, 15310 Athens, Greece\\

\bigskip

${}^c$ Departamento de F\'{i}sica Te\'orica y del Cosmos, CAPFE,
       Universidad de Granada, E-18071 Granada, Spain \\

~~~~~~~\\

 \email{  \hskip0.5cm} }

\abstract{We present a calculation of the NLO QCD corrections for the production 
of three vector bosons at the LHC, namely $Z Z Z$, $W^{+}W^{-}Z$, $W^{+} Z Z$, 
and $W^{+}W^{-}W^{+}$ production. The virtual corrections are computed using the 
recently proposed method of reduction at the integrand level (OPP reduction).
Concerning the contributions coming from real emission we used the dipole subtraction
to treat the soft and collinear divergences. We find that
the QCD corrections for these electroweak processes are in the 
range between 70 and 100 percent.  As such they have to be considered
in experimental studies of triple vector boson production at the LHC.
}


\begin{document}


\section{Introduction}

For TeV collider physics hard multi-particle final states are ubiquitous
and theoretical calculations can not provide reliable predictions without
taking into account higher order information.
Unfortunately the evaluation of one-loop amplitudes with many external
particles is technically very challenging, which motivated a priority list
for one-loop computations relevant for the Large Hadron Collider at CERN,
the so called Les Houches wish list \cite{Bern:2008ef}.
Due to the relevance for LHC phenomenology many new avenues have been
explored in the last few years, ranging from evaluation techniques of Feynman diagram
\cite{Giele:2004iy,Binoth:2005ff,Denner:2005nn,avh,Ellis:2005zh} to unitarity based approaches~\cite{Bern:2007dw} in  different variations
\cite{Britto:2006sj,Binoth:2007ca,Bernicot:2007hs,Ellis:2007br,Britto:2008vq,Giele:2008ve,Moretti:2008jj,Berger:2008sj}.

Higher order QCD results have been provided recently
for multi-boson production $pp\to ZZZ,WWZ,HHH$ processes
\cite{zzz,wwz,Plehn:2005nk,Binoth:2006ym},
in the context of weak boson fusion \cite{Jager:2006zc,Bozzi:2007ur,Andersen:2007mp,Ciccolini:2007ec,Bredenstein:2008tm},
$pp\to Hjj$ with  effective gluon-Higgs
couplings \cite{Campbell:2006xx},
$gg\to Hq\bar{q}$ \cite{Weber:2006au}, and $pp\to t\bar{t}j$ \cite{Dittmaier:2007wz}.

In two recent papers~\cite{Opp1, Opp2}, a new
technique (OPP) has been introduced for the reduction of arbitrary one-loop sub-amplitudes at {\it the
integrand level}~\cite{intlevel}
by exploiting numerically the set of kinematical
equations for the integration momentum, that extend the quadruple,
triple and double cuts used in the unitarity-cut method~\cite{Bern:1994cg,Britto:2004nc,unicut}.
The method requires a minimal information about the form of the one-loop
(sub-)amplitude and therefore it is well suited for a numerical
implementation.

In the present work, the OPP reduction is applied to the calculation of
the next-to-leading order QCD correction for the production
of three vector bosons at the LHC. This includes the case of $Z Z Z$ production,
as well as the $W^{+}W^{-}Z$, $W^{+} Z Z$, and $W^{+}W^{-}W^{+}$ production.
The physics motivation for a reliable prediction of these processes is two-fold:
firstly one is sensitive to quartic vector boson couplings and secondly the leptonic
decays are prominent Standard Model backgrounds for multi-lepton and missing energy
signatures present in many new physics scenarios.

As the triple vector boson production is genuinely an electroweak process
one can not expect that the inclusion of QCD effects leads to the reduced
scale dependence typically seen in this kind of calculations. In contrary
it can be qualitatively understood that  the LO predictions
show a relatively small sensitivity when varying the factorisation scale.
This is because the parton distribution functions are called for
x-values which are around the scaling region where one has a
very mild $Q^2$-dependence. After adding
the order $\alpha_s$ corrections one expects to observe
a LO type scale variation  in the added contribution.

The production of three $Z$ bosons has already been discussed by Lazopoulos et al. in Ref.~\cite{zzz}. We also presented some preliminary results in Refs.\cite{Bern:2008ef,zzztalks}.
The $W^{+}W^{-}Z$ case has been studied in
Ref.~\cite{wwz} for all combinations of leptonic final states. Results for $W^{+} Z Z$ and $W^{+}W^{-}W^{+}$ production have not been presented in the literature yet.

Our calculation is composed of two main parts: the evaluation of virtual corrections, namely one-loop contributions obtained adding a virtual particle to the tree-level diagrams, and corrections from the real emission of one additional massless particles from initial and final states, needed in order to control and cancel infrared singularities.
The virtual corrections are computed using the {\tt OPP} reduction~\cite{Opp1, Opp2}. In particular, we make
use of {\tt CutTools} \cite{cuttools}, a FORTRAN90 code that implements the general method of reduction.
Concerning the contributions
coming from real emission we used the dipole subtraction
method~\cite{Catani:1996vz} to isolate the soft and collinear
divergences and checked the results using the phase space slicing
method~\cite{Giele:1991vf,Giele:1993dj} with soft and collinear cutoffs, as
outlined in~\cite{Baur:1998kt,Harris:2001sx}.

The paper is organized as follows. In Section~\ref{virtual}, we report the details of the calculation of the virtual part. Section~\ref{real} is devoted to the discussion of soft and collinear singularities. In Section~\ref{numbers}, we show our results, including transverse momentum and rapidity distributions for the different processes studied in this paper. Finally, in Section~\ref{conclusion}, we will give a summary of the work done and present our conclusions.

\section{Virtual corrections} \label{virtual}

We consider the process
\begin{equation}
q (p_1) + {\bar q}(p_2) \longrightarrow V (p_3) + V (p_4)+ V (p_5)
\end{equation}
where $V={Z,W}$. All momenta are chosen to be incoming, such that $\sum_i p_i =0$.

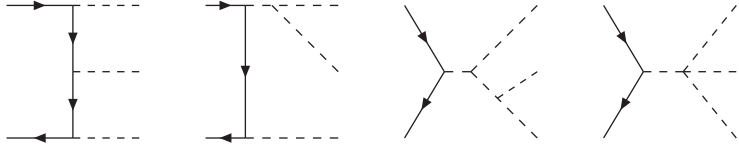
\begin{figure}[ht] \
     \begin{center}
      \begin{picture}(300,50)(0,0)
        \ArrowLine(0,50)(25,50)
        \ArrowLine(25,0)(0,0)
        \ArrowLine(25,25)(25,0)
    \ArrowLine(25,50)(25,25)
        \DashLine(25,0)(50,0){3}
        \DashLine(25,25)(50,25){3}
        \DashLine(25,50)(50,50){3}

  \ArrowLine(75,50)(90,50)
  \ArrowLine(90,0)(75,0)
  \ArrowLine(90,50)(90,0)
  \DashLine(90,50)(125,50){3}
  \DashLine(90,0)(125,0){3}
  \DashLine(100,50)(125,25){3}

    \ArrowLine(150,50)(165,25)
        \ArrowLine(165,25)(150,0)
        \DashLine(165,25)(175,25){3}
        \DashLine(175,25)(200,50){3}
        \DashLine(185,15)(200,25){3}
        \DashLine(200,0)(175,25){3}

    \ArrowLine(225,50)(240,25)
        \ArrowLine(240,25)(225,0)
        \DashLine(240,25)(255,25){3}
        \DashLine(255,25)(275,50){3}
        \DashLine(255,25)(275,25){3}
        \DashLine(275,0)(255,25){3}
    \end{picture}
     \end{center} \caption{Tree-level structures of Feynman diagrams contributing to  $q {\bar q} \to V V V$, where $V={Z,W}$. Dashed internal lines can represent W, Z, Goldstone bosons or photons.} \label{ftree}
\end{figure}

At the tree-level, diagrams can be grouped in four different topologies, which are
illustrated in Fig.~\ref{ftree}. One-loop corrections are obtained by adding a virtual gluon to the tree-level structures, as depicted in Figs.~\ref{fig1} and~\ref{fig2}.

We perform a reduction to scalar integrals using the {\tt OPP} reduction method~\cite{Opp1,Opp2}.
In this approach, we need to provide the numerical value of the numerator of the integrand in the loop integrals. We refer to it as the numerator function $N(q)$, where $q$ is the integration momentum.

The numerator function $N(q)$ can be expressed in terms of $4$-dimensional denominators $\d{i} = ({q} + p_i)^2-m_i^2$ as follows
\bqa \label{mastereq}
N(q) &=& \sum_{i_0 < i_1 < i_2 <
i_3}^{m-1} \left[
          d( i_0 i_1 i_2 i_3 ) +
     \tld{d}(q;i_0 i_1 i_2 i_3)
\right] \prod_{i \ne i_0, i_1, i_2, i_3}^{m-1} \d{i} \nl
     &+&
\sum_{i_0 < i_1 < i_2 }^{m-1} \left[
          c( i_0 i_1 i_2) +
     \tld{c}(q;i_0 i_1 i_2)
\right] \prod_{i \ne i_0, i_1, i_2}^{m-1} \d{i} \nl
     &+&
\sum_{i_0 < i_1 }^{m-1} \left[
          b(i_0 i_1) +
     \tld{b}(q;i_0 i_1)
\right] \prod_{i \ne i_0, i_1}^{m-1} \d{i} \nl
     &+&
\sum_{i_0}^{m-1} \left[
          a(i_0) +
     \tld{a}(q;i_0)
\right] \prod_{i \ne i_0}^{m-1} \d{i} \,. \eqa 
The quantities $d( i_0 i_1 i_2 i_3 )$ are the coefficients of 4-point
scalar functions with denominators labeled by $i_0$, $i_1$, $i_2$, and $i_3$. In the same way,  $c( i_0 i_1 i_2)$, $b( i_0 i_1)$, and $a( i_0)$ are the coefficients of the 3-point, 2-point and 1-point scalar functions, respectively.
The other quantities appearing in Eq.~(\ref{mastereq}), marked with a ``tilde'', vanish upon integration over $q$.
Such a separation is always possible and the set of coefficients $d, c, b, a$ is immediately interpretable as the
ensemble of the coefficients of all possible 4, 3, 2, 1-point one-loop functions contributing to the amplitude.

Since the structure of Eq.~(\ref{mastereq}) is general, namely independent from the
particular process that we want to study, the task of computing the one-loop amplitude
is then reduced to the algebraical problem of fitting the
coefficients $d,c,b,a$ by evaluating the function $N(q)$ a
sufficient number of times, at different values of $q$, and then
inverting the system. That can be achieved quite efficiently by
singling out particular choices of $q$ such that, systematically, 4,
3, 2 or 1 among all possible denominators $\d{i}$ vanishes.
 Then the system of equations is solved iteratively
\footnote{A method to optimize the solution of the system has been very recently presented in~\cite{Mastrolia:2008jb}.}.

First one determines all possible 4-point functions, then the
3-point functions and so on.
In summary, simply by evaluating the numerator function $N(q)$ for a given set of values of $q$, we can extract all the coefficients in Eq.~(\ref{mastereq}).

As a possible future development, the numerical evaluation of $N(q)$ could be performed automatically without relying on Feynman diagrams, by means of recursion relations. For the current project however, we still follow the traditional approach of computing all the expression originating from Feynman diagrams, and use them to evaluate numerically the numerator functions.
An example in Section~\ref{sub_zzz} will clarify the details of the technique used.

The coefficients determined in this manner should be multiplied by the corresponding scalar
integrals. Since, in the process that we are studying, no $q$-dependent massive propagators appear, we will only need massless scalar integrals. They are computed using the package {\tt OneLOop} written by A.~van~Hameren~\cite{avh}.

The last step is the calculation of the rational terms. As explained in Ref.~\cite{r1r2}, there are two sources
of the rational terms: a first contribution, that we call $R_1$, originates from considering the fact that the denominators appearing in the scalar integrals are n-dimensional objects, while the expansion of Eq.~(\ref{mastereq}) is purely 4-dimensional. These contributions can be automatically extracted in the reduction process, either by computing extra-integrals as explained in Ref.~\cite{Opp2}, or by means of a modified version of Eq.~(\ref{mastereq})
in which the numerator function is expressed directly in terms of n-dimensional denominators.
The second approach is illustrated in Ref.~\cite{r1r2} and implemented in
the package {\tt CutTools} \cite{cuttools}.
We checked that the results obtained for $R_1$ with the two methods are in perfect agreement.

The second contribution, that we call $R_2$, is instead originating from the numerator function. For many processes, $N(q)$ can be treated as a purely four-dimensional object. However, in general, it should be written as ${\bar N}(q) = N(q) + {\tilde N}(q^2,\epsilon)$,
where ${\bar N}(q)$ is the n-dimensional numerator function. ${\tilde N}(q^2,\epsilon)$ can originate, for example, from the contraction of Dirac matrices or from
powers of $q^2$ in the numerator function \cite{pittau2} and vanishes in the $\epsilon \to 0$ limit.
In Ref.~\cite{r1r2} we discussed in detail this topic and showed how $R_2$ can be obtained, for example, by using a set of tree-level like Feynman rules. For the calculation presented in this paper, however, it is easy to extract these remaining rational parts directly.

\subsection{ZZZ production} \label{sub_zzz}

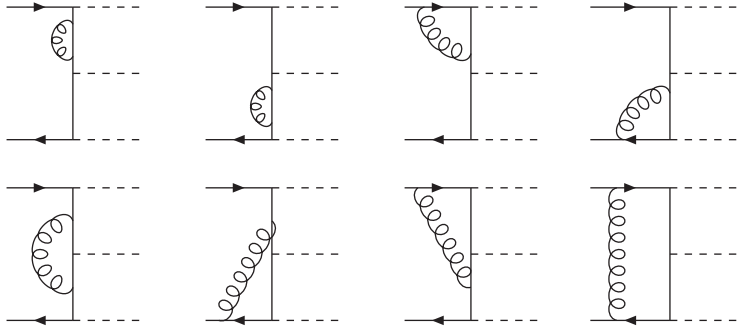
\begin{figure}[ht]
\begin{center} \begin{picture}(300,50)(0,0)
        \ArrowLine(0,50)(25,50)
        \ArrowLine(25,0)(0,0)
        \Line(25,0)(25,50)
        \DashLine(25,0)(50,0){3}
        \DashLine(25,25)(50,25){3}
        \DashLine(25,50)(50,50){3}
        \GlueArc(25,37.5)(6,90,270){2}{3}

        \ArrowLine(75,50)(100,50)
        \ArrowLine(100,0)(75,0)
        \Line(100,0)(100,50)
        \DashLine(100,50)(125,50){3}
        \DashLine(100,0)(125,0){3}
        \DashLine(100,25)(125,25){3}
    \GlueArc(100,12.5)(6,90,270){2}{3}

    \ArrowLine(150,50)(175,50)
        \ArrowLine(175,0)(150,0)
        \Line(175,0)(175,50)
        \DashLine(200,50)(175,50){3}
        \DashLine(200,0)(175,0){3}
        \DashLine(200,25)(175,25){3}
    \GlueArc(175,50)(17.5,180,270){3}{4}

        \ArrowLine(220,50)(250,50)
        \ArrowLine(250,0)(220,0)
        \Line(250,0)(250,50)
        \DashLine(250,50)(275,50){3}
        \DashLine(250,0)(275,0){3}
        \DashLine(250,25)(275,25){3}
    \GlueArc(250,0)(17.5,90,180){3}{4}
 \end{picture}
\begin{picture}(300,5)(0,0)
 \end{picture}
\begin{picture}(300,50)(0,0)
        \ArrowLine(0,50)(25,50)
        \ArrowLine(25,0)(0,0)
        \Line(25,0)(25,50)
        \DashLine(25,0)(50,0){3}
        \DashLine(25,25)(50,25){3}
        \DashLine(25,50)(50,50){3}
    \GlueArc(25,25)(12.5,90,270){3}{5}

        \ArrowLine(75,50)(100,50)
        \ArrowLine(100,0)(75,0)
        \Line(100,0)(100,50)
        \DashLine(100,50)(125,50){3}
        \DashLine(100,0)(125,0){3}
        \DashLine(100,25)(125,25){3}
        \Gluon(100,37.5)(80,0){3}{6}

    \ArrowLine(150,50)(175,50)
        \ArrowLine(175,0)(150,0)
        \Line(175,0)(175,50)
        \DashLine(200,50)(175,50){3}
        \DashLine(200,0)(175,0){3}
        \DashLine(200,25)(175,25){3}
        \Gluon(175,12.5)(155,50){3}{6}

        \ArrowLine(220,50)(250,50)
        \ArrowLine(250,0)(220,0)
        \Line(250,0)(250,50)
        \DashLine(250,50)(275,50){3}
        \DashLine(250,0)(275,0){3}
        \DashLine(250,25)(275,25){3}
        \Gluon(230,0)(230,50){3}{7}

      \end{picture}
     \end{center} \caption{Diagrams contributing to virtual QCD corrections
to $q {\bar q} \to Z Z Z$} \label{fig1}
\end{figure}

In this subsection we describe the evaluation of the virtual QCD corrections to the process $q {\bar q} \to Z Z Z$.
The virtual part of the calculation involves eight different diagrams, which have been depicted in Fig.~\ref{fig1}. Each diagram should be evaluated for six permutations of the final particles.

As an example, let us consider the pentagon diagram (the last diagram of Fig.~\ref{fig1}).
In our notation, the integrand will read
\begin{equation} \label{a5}
 A_5(q) = \frac{N_5(q)}{[q^2][(q+p_1)^2][(q+p_1+p_5)^2][(q-p_2-p_3)^2][(q-p_2)^2]}
\end{equation}
with
{\small
\begin{equation} \label{n5}
 N_5(q) = \left\{ {\bar u}(p_2)\,\gamma^\alpha\,P_{(q-p_2)} \, V^Z_3\, P_{(q-p_2-p_3)}\, V^Z_4\, P_{(q+p_1+p_5)}\, V^Z_5\, P_{(q+p_1)}\, \gamma^\alpha\, u(p_1) \right\}
\end{equation}
}

The function $P(q)$ is the numerator of the quark propagator
$$P_{(q)} = \rlap/ q \, , $$
while $V_i^Z = V^Z \cdot \epsilon_i$ ,
namely the contraction between the polarization vector of the \emph{i-th} $Z$ boson $\epsilon_i$
and the $\gamma$-matrix in the vertex $Zq{\bar q}$
\begin{equation}
 V^Z_\mu = ie\gamma_{\mu}( g_{f}^{-}  \omega_{-} + g_{f}^{+} \omega_{+})
\end{equation}
where
\begin{equation}
 g_{f}^{+} = -\frac{s}{c} Q_{f} \,\,\, , \,\,\,  g_{f}^{-} = \frac{I_{W,f}^{3}-s^{2}Q_{f}}{sc} \,\,\,
, \,\,\,  \omega_{\pm} = (1 \pm \gamma^5)/2\,,\, s=\sin\theta_W\,,\, c=\cos\theta_W.
\end{equation}
For any fixed value $q_0$ of integration momentum, and for a given phase space point,
$N_5(q_0)$ is simply the trace of a string of known matrices. After choosing a representation for Dirac matrices and spinors, we evaluate $N(q)$ by performing a naive matrix multiplication. By providing this input to the reduction algorithm, we can compute all the coefficients of the scalar integrals (in other words, the ``cut-constructible'' part of the calculation).

In the same fashion, we can repeat the calculation for the other seven diagrams. However, our method allows for a further simplification: for each fixed permutation of the final legs, only the q-dependent denominators of Eq.~(\ref{a5}) will appear also in the remaining diagrams. Therefore, we can combine all diagrams in a single numerator function and perform the reduction directly for the sum of such diagrams, allowing for a one-shot evaluation of the resulting scalar coefficients.

We checked that our results, both for poles and finite parts, agree with the results obtained by the authors of Ref.~\cite{zzz}.

\subsection{$W^{+}W^{-}Z$ production}

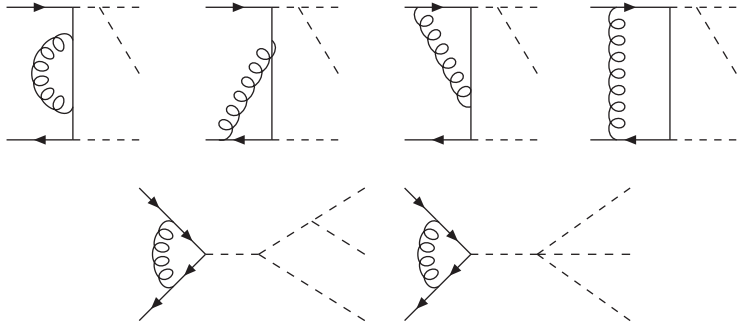
\begin{figure}[htb] \
     \begin{center}
      \begin{picture}(300,50)(0,0)
         \ArrowLine(0,50)(25,50)
        \ArrowLine(25,0)(0,0)
        \Line(25,0)(25,50)
        \DashLine(25,0)(50,0){3}
        \DashLine(35,50)(50,25){3}
        \DashLine(25,50)(50,50){3}
    \GlueArc(25,25)(12.5,90,270){3}{6}

        \ArrowLine(75,50)(100,50)
        \ArrowLine(100,0)(75,0)
        \Line(100,0)(100,50)
        \DashLine(100,50)(125,50){3}
        \DashLine(100,0)(125,0){3}
        \DashLine(110,50)(125,25){3}
        \Gluon(100,37.5)(80,0){3}{6}

    \ArrowLine(150,50)(175,50)
        \ArrowLine(175,0)(150,0)
        \Line(175,0)(175,50)
        \DashLine(200,50)(175,50){3}
        \DashLine(200,0)(175,0){3}
        \DashLine(185,50)(200,25){3}
        \Gluon(175,12.5)(155,50){3}{6}

        \ArrowLine(220,50)(250,50)
        \ArrowLine(250,0)(220,0)
        \Line(250,0)(250,50)
        \DashLine(250,50)(275,50){3}
        \DashLine(250,0)(275,0){3}
        \DashLine(260,50)(275,25){3}
        \Gluon(230,0)(230,50){3}{7}

    \end{picture}
\begin{picture}(300,5)(0,0)
 \end{picture}
 \begin{picture}(300,50)(0,0)

    \ArrowLine(50,50)(62,38)
        \ArrowLine(62,38)(75,25)
        \ArrowLine(75,25)(62,12)
    \ArrowLine(62,12)(50,0)
    \GlueArc(75,25)(17,135,225){3}{4}
        \DashLine(75,25)(95,25){3}
        \DashLine(95,25)(135,50){3}
        \DashLine(115,37.5)(135,25){3}
        \DashLine(95,25)(135,0){3}

    \ArrowLine(150,50)(162,38)
        \ArrowLine(162,38)(175,25)
        \ArrowLine(175,25)(162,12)
    \ArrowLine(162,12)(150,0)
    \GlueArc(175,25)(17,135,225){3}{4}
        \DashLine(175,25)(200,25){3}
        \DashLine(200,25)(235,50){3}
        \DashLine(200,25)(235,25){3}
        \DashLine(200,25)(235,0){3}
    \end{picture}
     \end{center} \caption{Additional NLO structures contributing in the production of $W^{+}W^{-}Z$, $W^{+} Z Z$, and $W^{+}W^{-}W^{+}$, that do not appear in the $Z Z Z$ case. Dashed internal lines can represent W, Z, Goldstone bosons or photons.} \label{fig2}
\end{figure}

With the same technique we also evaluated the virtual QCD corrections to the process $q {\bar q} \to W^{+}W^{-}Z$. The structure of the diagrams is more involved with respect to the $ZZZ$ case. There are in fact 19 different tree level diagrams. Adding QCD corrections, we obtain 58 one-loop diagrams contributing to this process. In addition to the structures already depicted in Fig.~\ref{fig1}, in this case we can also have the diagrams appearing in Fig.~\ref{fig2}.

A very similar calculation has been presented recently by Hankele and Zeppenfeld \cite{wwz}. They studied
the NLO QCD corrections to the production of 6 leptons in hadronic collisions, via  $W W Z$ production. A comparison with their results, however, is not straightforward and has not been performed yet.

\subsection{$W^{+} Z Z$ production}

Concerning the production of $W^{+} Z Z$, we have
15 tree-level diagrams, which, after adding QCD corrections, give rise to 69 diagrams at NLO. Since in this process we generate a single charged $W^{+}$, the initial
state should be of the type $u {\bar d}$ (rather then $u {\bar u}$ as for $W^{+}W^{-}Z$ and $Z Z Z$).

\subsection{$W^{+}W^{-}W^{+}$ production}
Starting again from an initial state of the type $u {\bar d}$, we should consider 15 diagrams at the tree-level and 53 diagrams including NLO QCD corrections.

\section{Real emission} \label{real}

The real emission corrections for the production process of three vector bosons
\bea\label{born}
q + \bar{q} \to V + V + V
\eea
fall in the following three categories
\bea
q + \bar{q} &\to& V + V + V + g  \label{ini_qq}\\
g +      q  &\to& V + V + V + q  \label{ini_gq}\\
g +\bar{q}  &\to& V + V + V + \bar{q}  \label{ini_gqb}\quad .
\eea
IR divergences arise if a massless final state particle becomes
soft or collinear to an initial parton. We deal with  the
IR part of the calculation by using the two cut-off phase space slicing
method \cite{Baur:1998kt,Harris:2001sx} and the dipole formalism of
Catani and Seymour \cite{Catani:1996vz}. Let us first provide the relevant formulas
for the dipole subtraction method.

\subsection{Dipole subtraction}

The partonic  cross section at the NLO level
consists of Born term $(B)$, virtual corrections $(V)$, collinear counter
terms $(C)$
defined on the 3-particle phase space and the real emission corrections $(R)$.
Dipole terms $(A)$ which approximate the real emission matrix elements
in all soft/collinear  regions are subtracted
from the real matrix element before integration over the four-particle phase space.
The same terms are added back,  integrated  over the dimensionally regulated 
phase space of the soft/collinear particle:
\bea
\sigma_{q\bar{q}}^{NLO} &=& \int\limits_{VVVg} \Bigl[ d\sigma_{q\bar{q}}^R
- d\sigma_{q\bar{q}}^A\Bigr]  + \int\limits_{VVV}\Bigl[ d\sigma_{q\bar{q}}^B + d\sigma_{q\bar{q}}^V
+ \int\limits_{g} d\sigma_{q\bar{q}}^A + d\sigma_{q\bar{q}}^C\Bigr] \quad .
\eea
After subtracting the dipole terms the real emission cross section is finite and can be
evaluated in 4 dimensions. The same is true for the other terms after the pole parts
have been canceled.

The colour averaged leading order contribution is given by
\bea
d\sigma_{q\bar{q}}^B = \frac{C_S}{N} \frac{1}{2 s_{12}}\, | \mathcal{M}^B|^2\; d\Phi_{VVV}
\eea
where $\mathcal{M}^B$ is the kinematic part of the leading order amplitude and $s_{12}=2 p_1 \cdot p_2$.
If two (three) vector bosons are identical a symmetry factor $C_S=1/2$ $(C_S=1/6)$ has
to be included. The three particle phase space of the vector bosons is
denoted as $d\Phi_{VVV}$.
The real emission corrections are defined on the four particle phase space $d\Phi_{VVVx}$
where $x$ can be either $g$, $q$, or $\bar{q}$.

In the case of a $q\bar{q}$ initial state two dipoles are needed as subtraction terms.
The subtraction term for the gluon emission off the quark
(neglecting $\mathcal{O}(\epsilon)$ terms) is
\bea
\mathcal{D}^{q_1g_6,\bar{q}_2} &=&
  \frac{8\pi \alpha_s C_F}{2 \tilde{x}\,p_1\cdot p_6}
  \left( \frac{1+\tilde{x}^2}{1-\tilde{x}} \right)
   |\mathcal{M}_{q\bar{q}}^B(\tilde{p}_{16},p_2,\tilde{p}_{3},\tilde{p}_{4},\tilde{p}_{5})|^2
\eea
where
\bea\label{dp_kinematics}
\tilde{x} &=& \frac{p_1\cdot p_2 - p_2\cdot p_6 - p_1\cdot p_6}{p_1\cdot p_2} \nl
\tilde{p}_{16} &=& \tilde{x}\, p_1 \quad , \quad K = p_1+p_2 -p_6  \quad , \quad \tilde{K}
= \tilde{p}_{16} + p_2     \nl
\Lambda^{\mu\nu} &=& g^{\mu\nu} -
\frac{2  ( K^\mu+\tilde{K}^\mu ) (K^\nu + \tilde{K}^\nu )}{(K+\tilde{K})^2}
             + \frac{2  \tilde{K}^\mu K^\nu}{K^2}  \nl
\tilde{p}_j &=& \Lambda \; p_j
\eea
defines the dipole kinematics: $q(\tilde{p}_{16}) + \bar{q}(p_2) \to V(\tilde{p}_3)
+ V(\tilde{p}_4) + V(\tilde{p}_5)$.
The subtraction term for gluon emission off the anti-quark is obtained by 
interchanging  $p_1 \leftrightarrow p_2$.
The real emission cross section including subtraction terms
reads
\bea
  d\sigma_{q\bar{q}}^R - d\sigma_{q\bar{q}}^A &=&
\frac{C_S}{N} \frac{1}{2 s_{12}} \Bigl[ C_F \, | \mathcal{M}_{q\bar{q}}^R|^2
 - \mathcal{D}^{q_1g_6,\bar{q}_2}  - \mathcal{D}^{\bar{q}_2g_6,q_1}  \Bigr] d\Phi_{VVVg}
\eea
The part of the NLO cross section which is defined on the $2\to3$ phase space is obtained after analytic
integration of the dipole terms over the phase space of the unresolved particle.
A collinear counter term is added to treat the collinear $1/\epsilon$ pole which is
absorbed into the parton distribution functions at a scale $\mu_F$. 
All details can be found in \cite{Catani:1996vz}.
The part which has to be added to the virtual corrections is given by
\bea
\label{virtsub}
&& d\sigma_{q\bar{q}}^C  + \int\limits_{g} d\sigma_{q\bar{q}}^A  =
\frac{\alpha_s C_F}{2\pi}\frac{\Gamma(1+\epsilon)}{(4\pi)^{-\epsilon}}
\left( \frac{s_{12}}{\mu^2}\right)^{-\epsilon} \Bigl[ \frac{2}{\epsilon^2} + \frac{3}{\epsilon}
- \frac{2\pi^2}{3}
\Bigr]d\sigma^B_{q\bar{q}} \nonumber
\\&&\quad
+ \frac{\alpha_s C_F}{2\pi} \int\limits_0^1 dx\;
\mathcal{K}^{q,q}(x) \, d\sigma^B_{q\bar{q}}(x p_1,p_2)
+ \frac{\alpha_s C_F}{2\pi} \int\limits_0^1 dx\;\mathcal{K}^{\bar{q},\bar{q}}(x) \,
d\sigma^B_{q\bar{q}}(p_1,x p_2)
\eea
where the term
\bea
\mathcal{K}^{q,q}(x) = \mathcal{K}^{\bar{q},\bar{q}}(x) &=&
\left[ \frac{1+x^2}{1-x}\right]_+ \log\left(\frac{s_{12}}{\mu_F^2} \right)
      + \left[ \frac{4\log(1-x)}{1-x}\right]_+       + (1-x) - 2 (1+x)\log(1-x)\non
\eea
contains plus distributions which are defined as usual
\bea
\int \limits_{0}^1 dx \, \Bigl[\frac{g(x)}{1-x}\Bigr]_+ f(x) = \int \limits_{0}^1 dx \, g(x) \frac{ f(x) - f(1)}{1-x}
\eea
For initial states with a gluon no soft contribution is present and thus one has
\bea
\sigma_{gq}^{NLO} &=&  \int\limits_{VVV}
\Bigl[  \int\limits_{q} d\sigma_{gq}^A + d\sigma_{gq}^C \Bigr]
+ \int\limits_{VVVq} \Bigl[ d\sigma_{gq}^R - d\sigma_{gq}^A \Bigr]
\eea
In this case only one subtraction term is needed, namely
 \bea
  d\sigma_{gq}^R - d\sigma_{gq}^A &=&
\frac{C_S}{N} \frac{1}{2 s_{12}} \Bigl[ T_R | \mathcal{M}_{gq}^R|^2
 - \mathcal{D}^{g_1q_6,q_2}  \Bigr] d\Phi_{VVVq} \quad ,
\eea
where the dipole is given by
\bea
\mathcal{D}^{g_1q_6,q_2} &=&
  \frac{8\pi \alpha_s \, T_R}{\tilde{x} \;2\,p_1\cdot p_6} \;  [ 1 - 2 \,\tilde{x}\, (1-\tilde{x}) ]
   \; |\mathcal{M}_{q\bar{q}}^B({\tilde{p}_j})|^2 \quad .
\eea
The momentum mappings ${\tilde{p}_j}$ are identical to the ones in Eq. (\ref{dp_kinematics}).

The initial state collinear singularity is again 
absorbed by the pdfs through a counter term
\bea
&& d\sigma^C_{gq} + \int\limits_{q} d\sigma^{A}_{gq} =
 \frac{\alpha_s T_R}{2\pi} \int\limits_0^1 dx \, \mathcal{K}^{g,q}(x)\,
d\sigma^B_{q\bar{q}}(xp_1, p_2)   \nl &&
\mathcal{K}^{g,q}(x) = [ x^2 + (1-x)^2 ] \log\left(\frac{s_{12}}{\mu_F^2} \right)
                + 2 x (1-x) + 2 [x^2 + (1-x)^2]\log(1-x) \quad .
\eea
The formulas for the cases $qg$, $\bar{q}g$, $g\bar{q}$ are identical up to 
relabeling of momenta.

The hadronic differential cross section with hadron momenta $P_1$ and $P_2$
is the sum over all partonic initial states
convoluted with the parton distribution functions
\bea
d\sigma(P_1,P_2) = \sum\limits_{ab} \int dz_1 dz_2 f_a(z_1,\mu_F) f_b(z_2,\mu_F) d\sigma_{ab}(z_1P_1,z_2P_2)
\eea
The sum runs over the six partonic configurations $q\bar{q}$, $\bar{q}q$, $gq$, $qg$, $g\bar{q}$, $\bar{q}g$.

\subsection{Phase space slicing}

To have an independent check for the real radiation 
we have also implemented the phase space
slicing method in its two cut-off variant~\cite{Baur:1998kt,Harris:2001sx}.
One splits the phase space in soft, collinear and hard regions
with the help of the cut-off parameters $\delta_s$ and $\delta_c$.
In the soft region the $2\to 4$ matrix element is replaced by the eikonal
approximation. In the collinear region  one has 
a convolution of a splitting function with the Born term.  Adding the soft/collinear
parts to the virtual corrections all poles
cancel and one obtains the three-particle contribution
\bea \sigma^{(3)} &=&\left(
\frac{\alpha_s}{2\pi} \right) \sum\limits_{a,b}\int dz_1 dz_2 d \sigma^B_{ab} \left[
f_{a}(z_1,\mu_F) f_{b}(z_2,\mu_F)
\left(A_0^s+A_0^v+2A_0^{\rm sc}\right) \right. \nl
&+& \left. f_{a}(z_1,\mu_F)
\widetilde{f}_{b}(z_2,\mu_F) +
\widetilde{f}_{a}(z_1,\mu_F) f_{b}(z_2,\mu_F) \right] \,
\eea
with
\bea
A_0^s &=&4 \ln^2\delta_s \,C_F\, \nl
A_0^{sc}(q\to qg) &=& C_F\,\left( 2 \ln \delta_s + 3/2 \right)
\ln\frac{s_{12}}{\mu^2_f} \nl
A_0^v &=& \frac{d \sigma^V_{ab}}{d \sigma^B_{ab}}\;.
\eea
The $\widetilde{f}$ functions are given by
\bea
\widetilde{f}_{a}(x,\mu_F) = \sum_{b}  \int_x^{1-\delta_s\delta_{ab}} \frac{dz}{z}
                       f_{b}(x/z,\mu_F) \widetilde{P}_{ab}(z)\,\, .
\label{eqn:g_tilde}
\eea
where
\bea
\widetilde{P}_{ab}(z) = P_{ab}(z)\ln\left(\delta_c\,\frac{1-z}{z}\,
\frac{2 x p_1\cdot p_2}{\mu_F^2}\right) - P_{ab}^{\prime}(z) \, .
\eea
The upper limit $1-\delta_s$ ensures that the soft region which is
already dealt with is excluded. The Kronecker  $\delta_{ab}$
indicates that for $a\neq b$ there is only a collinear
divergence and no soft cut-off is needed.

In our case we need only the splitting functions
$P_{qq}(z)$ and $P_{gq}(z)$.
If we write $P_{ab}(z,\epsilon)=P_{ab}(z)+\epsilon P^{\prime}_{ab}(z)$, we have
\bea
P_{qq}(z) &=& C_F \frac{1+z^2}{1-z} \label{eqn:ap_unreg1} \\
P_{qq}^{\prime}(z) &=& -C_F(1-z)\nonumber \\
P_{gq}(z) &=& T_R\,(z^2+(1-z)^2) \nonumber \\
P_{gq}^{\prime}(z) &=& -2\,T_R\,z(1-z) \nonumber
\eea
We see that the $\widetilde{f}$ functions contain an
explicit logarithm of $\delta_c$ as well as logarithmic dependencies on
$\delta_s$ which are built up by the integration on $z_1,z_2$.

The four-body contribution is given by \bea
\sigma^{(4)} = \sum_{a,b=\bar{q},q,g} \int dz_1 dz_2 f_{a}(z_1,\mu_F)
f_{b}(z_2,\mu_F) d\hat{\sigma}_{ab}^{R} \, , \eea with the
hard-non-collinear partonic cross section given by \bea
d\hat{\sigma}_{ab} = \frac{C_S}{2s_{12}}\int_{H\overline C} \overline{\sum}
|\mathcal{M}_{ab}|^2 d\Phi_{VVVx} \, , \eea where
$\overline{\sum} |\mathcal{M}^R|^2$ is the two-to-four body squared
matrix element averaged (summed)
over initial (final) degrees of freedom, $d\Phi_{VVVx}$
is the four-body phase space and the hard non-collinear region denoted by
$H\overline C$ is defined by \bea
E_6 &>& \delta_s\,\frac{\sqrt{s_{12}}}{2}\nonumber\\
2 p_1\cdot p_6,2 p_2\cdot p_6&>&\delta_c\,s_{12} \eea
where $p_6$ is the momentum of the soft/collinear parton with energy $E_6$.

Both methods have been implemented in {\tt HELAC}\cite{helac}.
The results show excellent agreement between the two methods. In the
numerical results presented below we only show the results of
the dipole subtraction approach.

\section{Numerical results} \label{numbers}

We present in this Section a selection of the results that we obtained
for the four processes studied in this paper.

The complete virtual part of the next-to-leading order calculation for the four processes studied
in this paper has been performed using {\tt CutTools} \cite{cuttools}
and also checked against an independent code.
The two programs provide identical results for the amplitudes studied. 
As further tests, we checked that the tree-level results obtained using Feynman
diagrams coincides with the results obtained with HELAC\cite{helac}
and that we reconstruct the correct structure for the poles after integration.
Concerning the finite parts, we agree with the results obtained by the authors of Ref.~\cite{zzz},
for the production of three $Z$ bosons.
In this section we will mostly focus on the processes for which no results have appeared
yet in the literature, namely the production of $W^{+}W^{-}W^{+}$  and $W^{+} Z Z$.

We use the following values for the electroweak parameters:
\begin{equation}
M_W=80.4\,\, \rm{GeV} \,\, , \,\, M_Z=91.1875\,\, \rm{GeV} \,\, , \,\,
G_F=1.16639 \cdot{10}^{-5} \,\, {\rm{GeV}}^{-2} \,\, .
\end{equation}
In all cases presented here, we set $\sqrt{s}=14$ TeV and used
CTEQ6L1~\cite{Pumplin:2002vw} with  $\alpha_s(M_Z)=0.129$ at NLO. For
the electroweak couplings we use the the $G_\mu$ scheme with
\begin{equation}\label{alf}
    \alpha_{em} = \frac{\sqrt{2}\, G_F M_W^2\sin^{2}\theta_W}{\pi}
\end{equation}
and
\begin{equation}\label{sin}
    \sin^{2}\theta_W=1-M_W^2/M_Z^2
\end{equation}

The tree-level cross section has been evaluated using the {\tt
HELAC} event generator\cite{helac}. The same programme,
appropriately adapted, has been used to calculate also the real
corrections. The virtual corrections have been calculated on the
basis of unweighted tree-order events produced by {\tt HELAC} with
an indicative CPU time of 180 ms per event, which is quite good
taking into account that the numerical calculation of one-loop
amplitudes (the numerators of the OPP method) is performed using
standard momentum representation of Feynman graphs without any
optimization. A conservative comparison with the efficiency of the
tree order calculation, based on {\tt HELAC}, shows that a further
improvement of the order of $10^1-10^2$ is to be expected.

Since the purpose of our paper is to show  the feasibility of the
OPP method in a realistic situation, the results we present here are
indicative and they constitute by no means a detailed discussion of
the phenomenology of these processes. Partial results have been
already presented in \cite{Bern:2008ef,zzztalks}.

It should be mentioned however that all results are available as
(un)weighted events, which means that an exhaustive study in the
full phase space, both for three- and four-particle final
states\footnote{Of course both positive and negative contributions
have been taken into account, separately.}, poses no problem and
will be postponed for the future, taking into account also decay
products and intermediate Higgs contributions.

\begin{figure}[ht]
\begin{center}
\begin{tabular}{cc}
  \begin{rotate}{90}\hspace*{35mm}{\tiny
  $d\sigma/ d p_T$ [pb/GeV]} \end{rotate}
  \psfig{figure=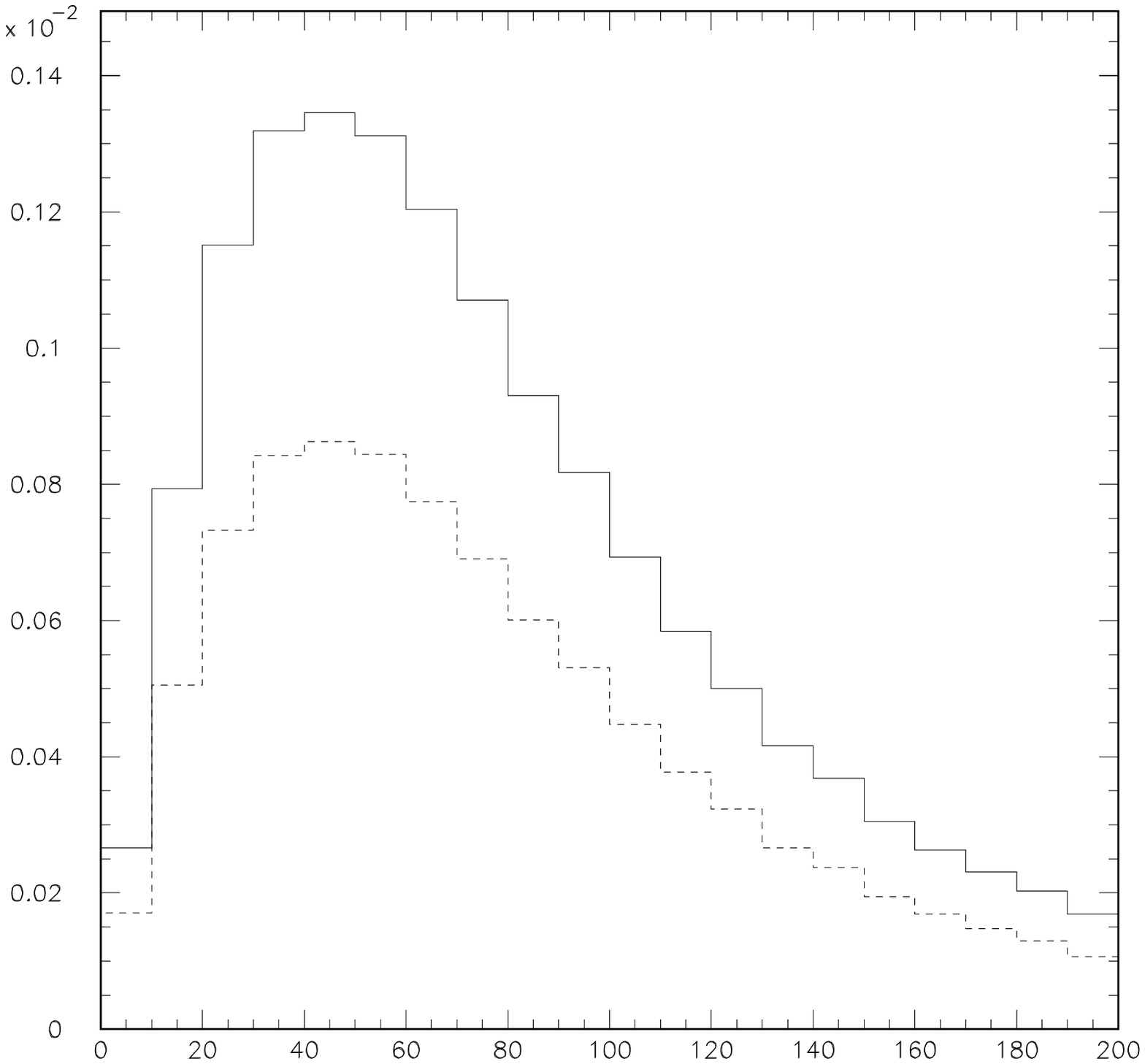,width=3.2truein} &
\begin{rotate}{90}\hspace*{35mm}{\tiny
  $d\sigma/ d p_T$ [pb/GeV]} \end{rotate}
  \psfig{figure=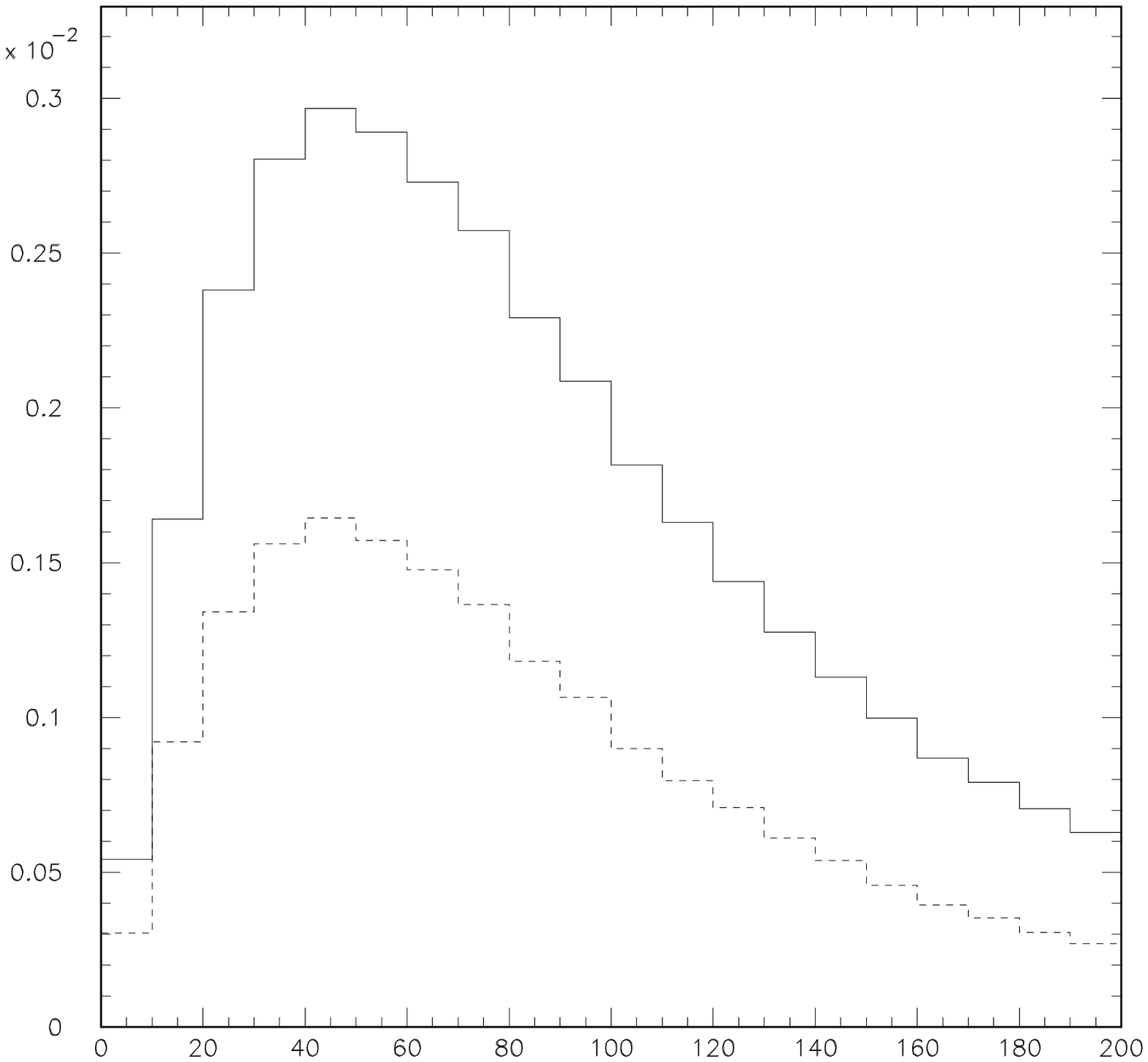,width=3.2truein} \\
\small{$Z Z Z$} & \small{$W^+ Z Z$} \\
\begin{rotate}{90}\hspace*{35mm}{\tiny
  $d\sigma/ d p_T$ [pb/GeV]} \end{rotate}
  \psfig{figure=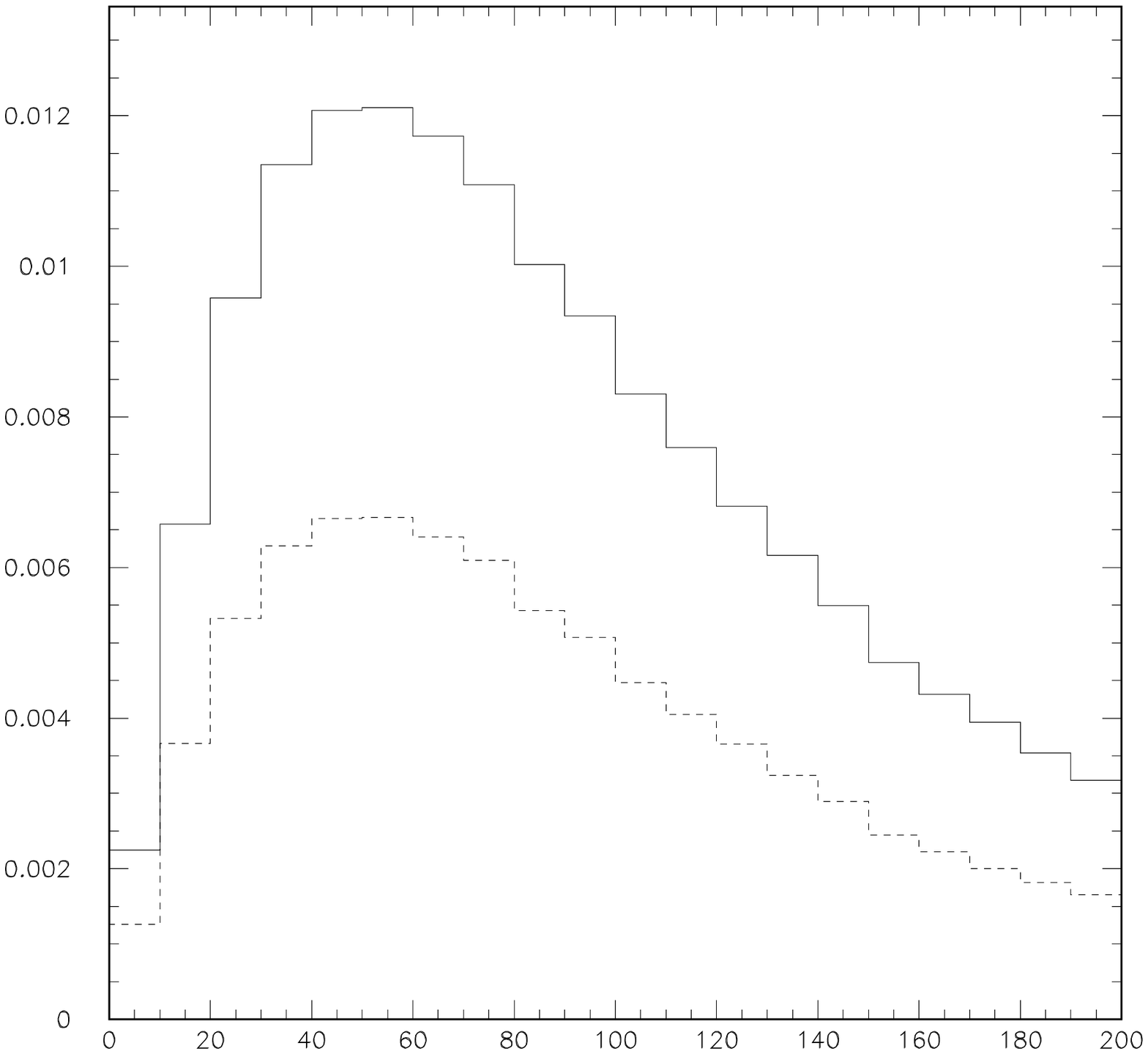,width=3.2truein} &
\begin{rotate}{90}\hspace*{35mm}{\tiny
  $d\sigma/ d p_T$ [pb/GeV]} \end{rotate}
  \psfig{figure=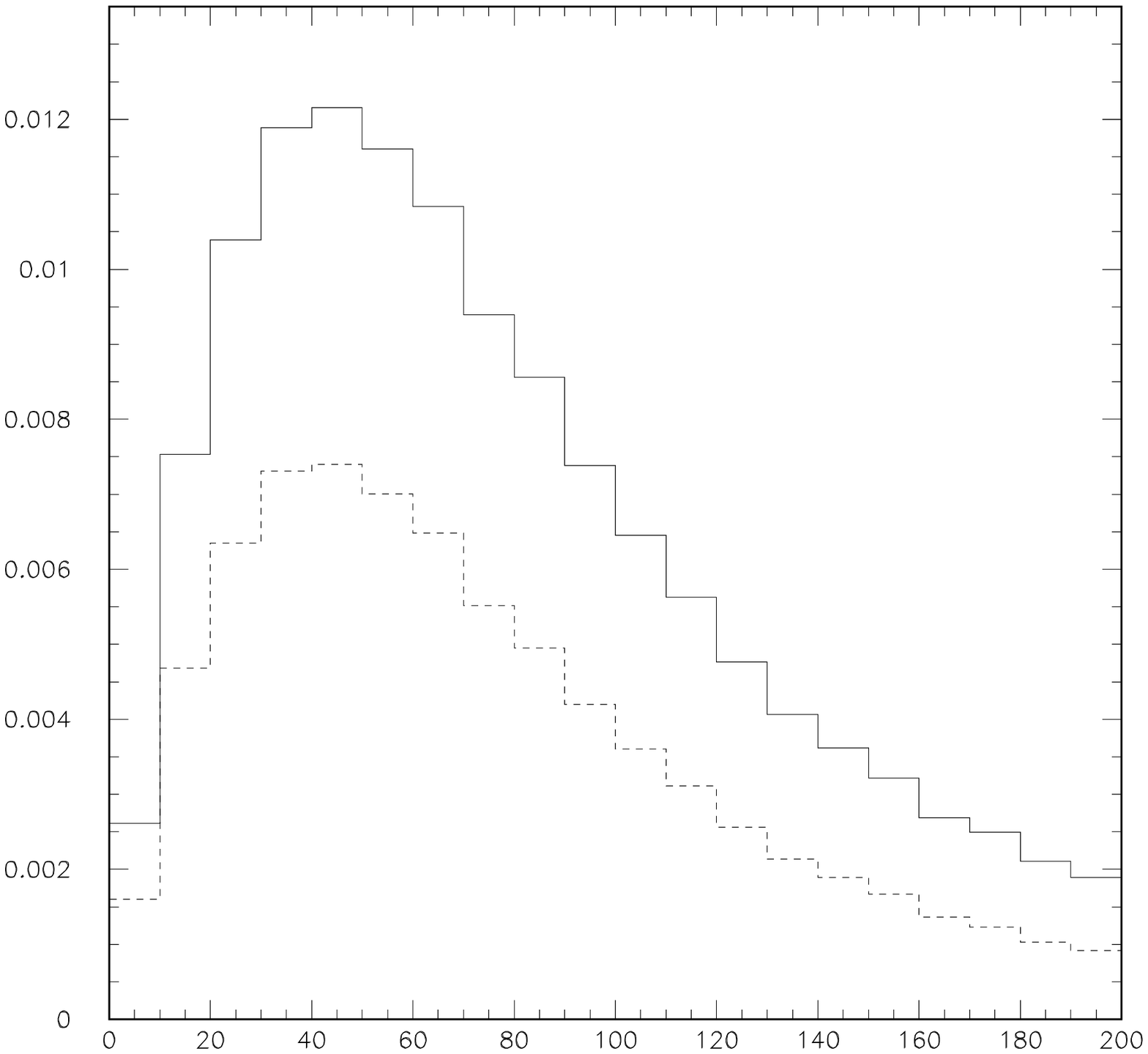,width=3.2truein} \\
\small{$W^{+} W^{-} Z$} & \small{$W^{+}W^{-}W^{+}$} \\
\end{tabular}
 \caption{Transverse momentum distribution, as defined in the text,  for the four processes $pp \to VVV$: NLO (solid line)
 compared with the LO contribution (dashed line). }  \label{ptfig}
 \end{center}
 \end{figure}

\begin{figure}[ht]
\begin{center}
\begin{tabular}{cc}
  \psfig{figure=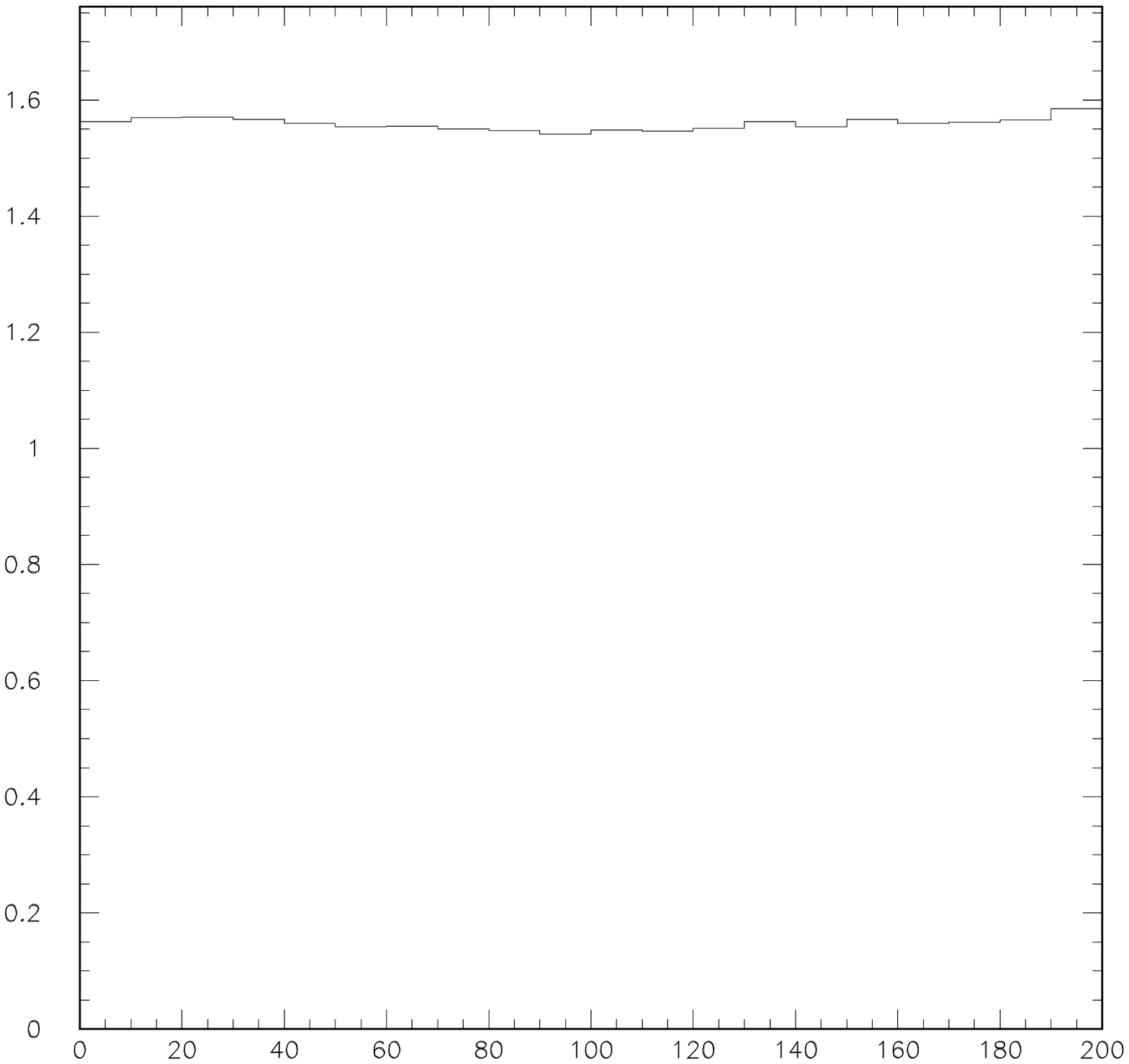,width=2.8truein} &
   \psfig{figure=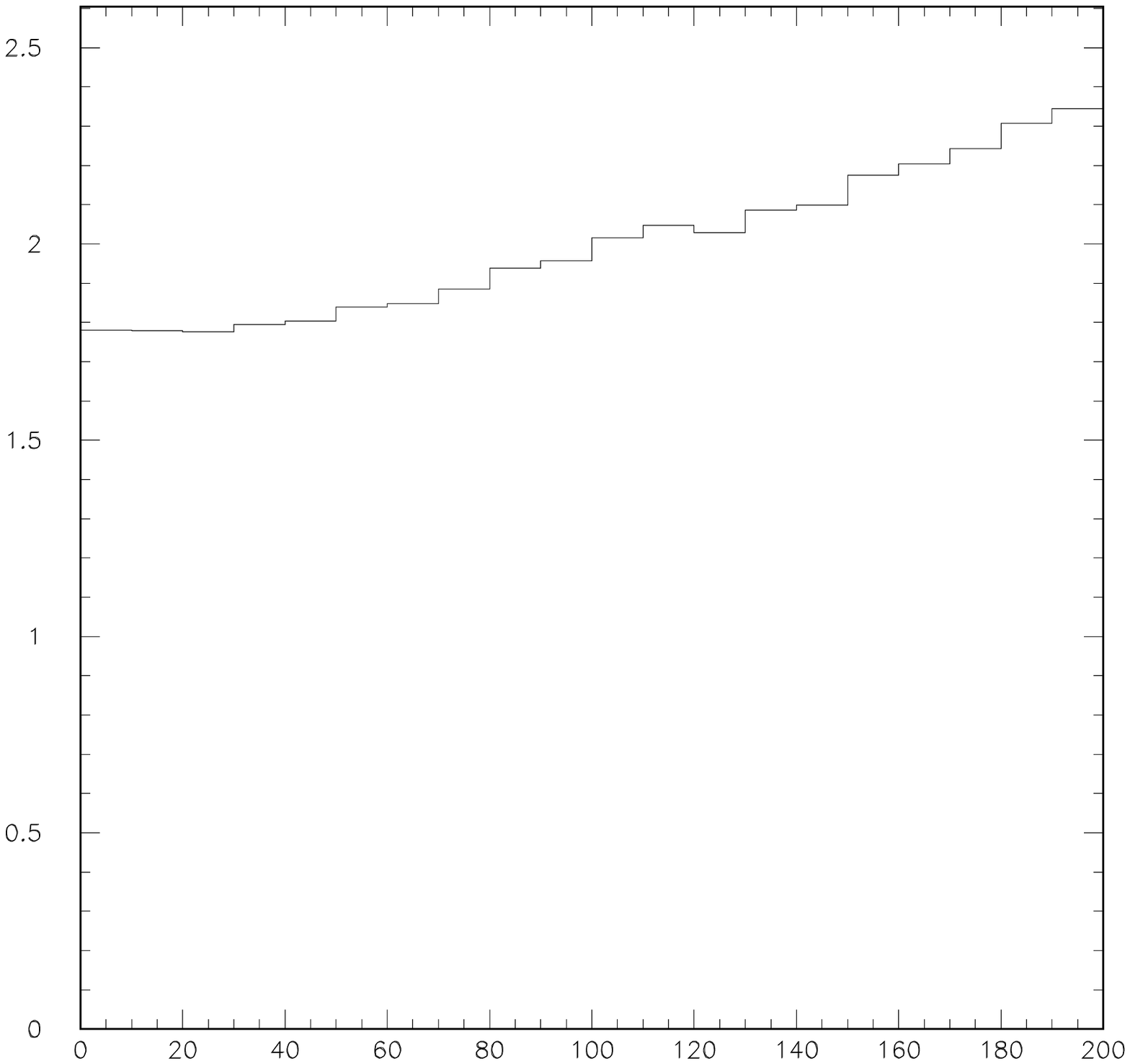,width=2.8truein} \\
\small{$Z Z Z$} & \small{$W^+ Z Z$} \\
   \psfig{figure=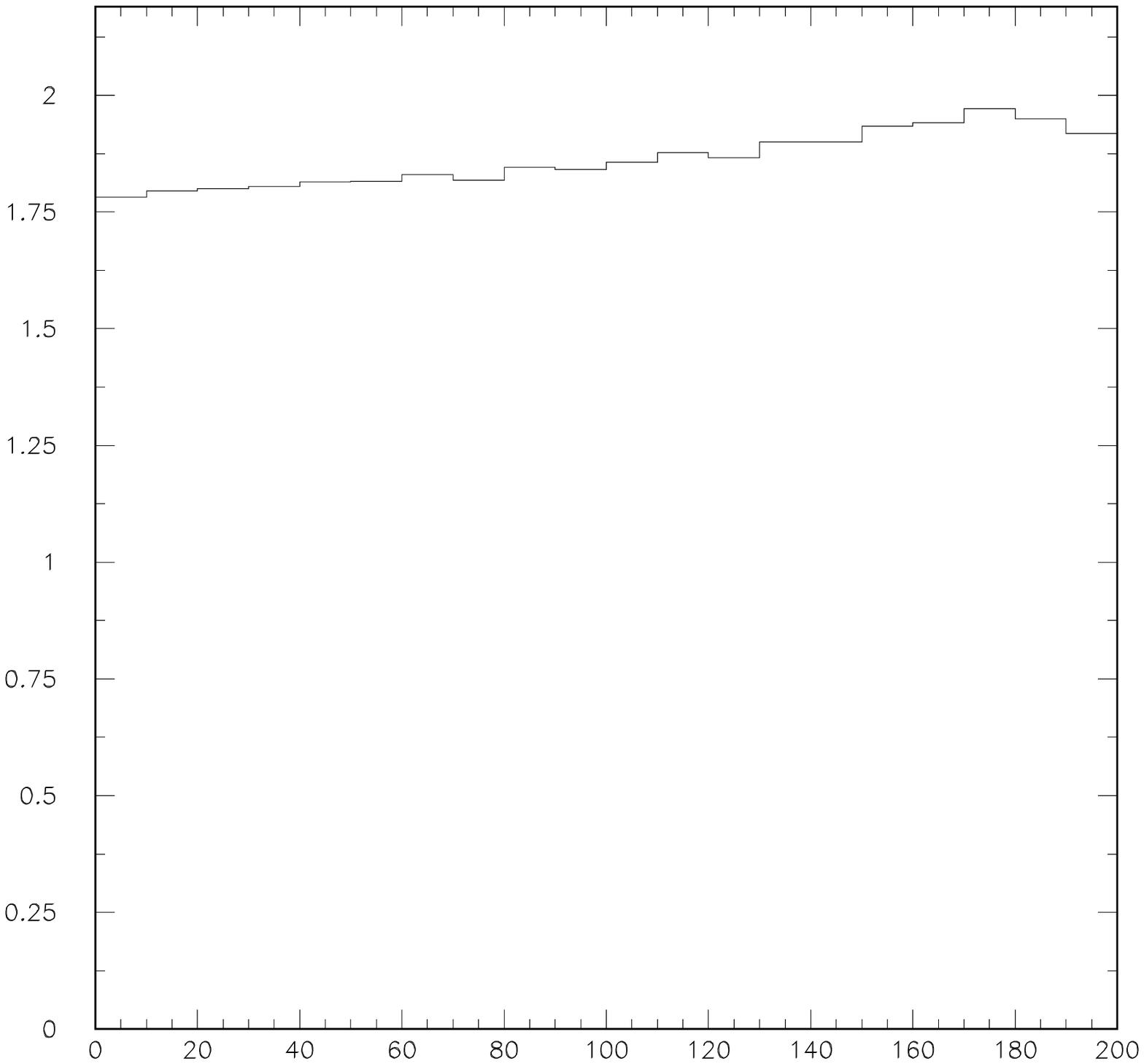,width=2.8truein} &
   \psfig{figure=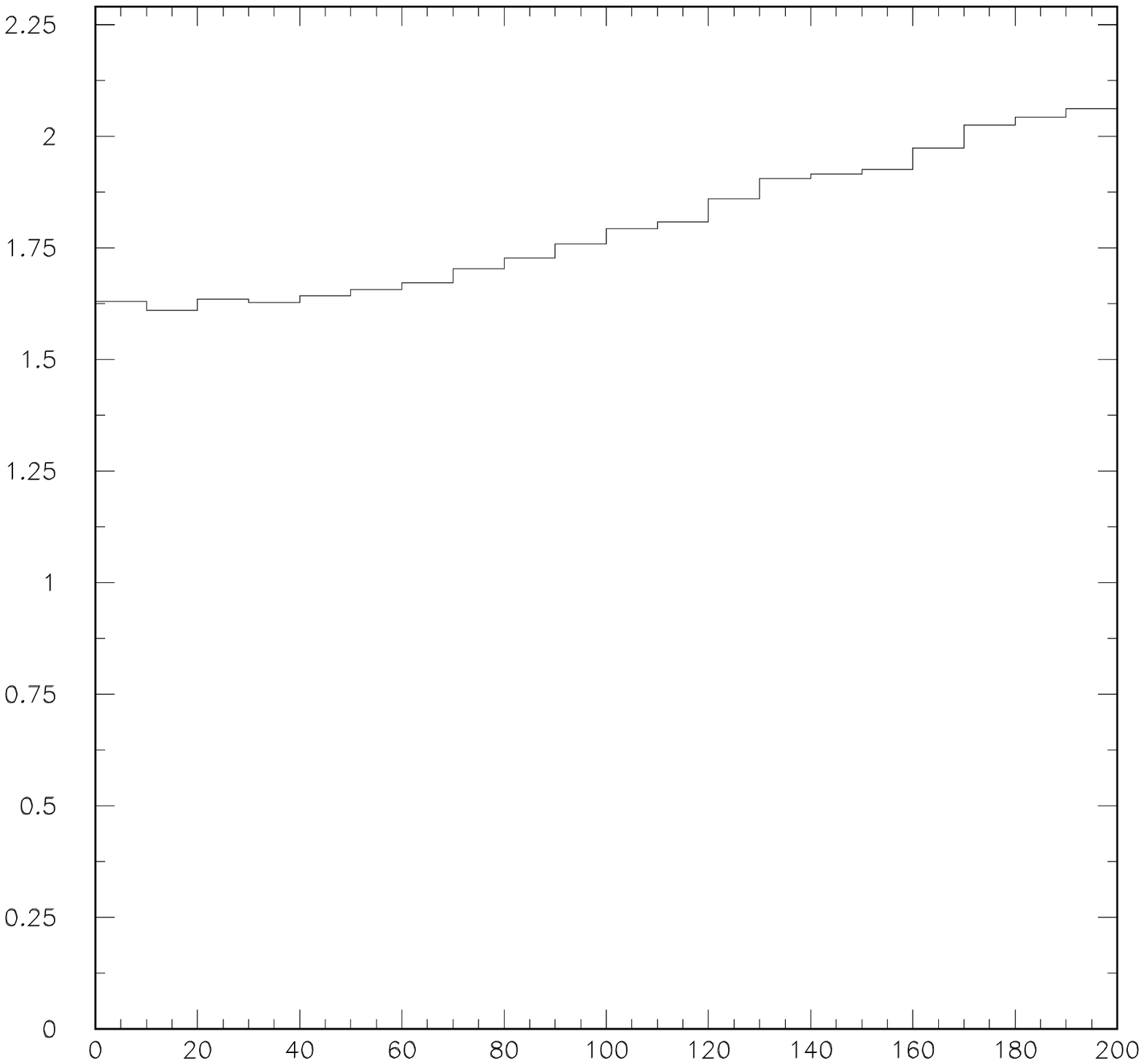,width=2.8truein} \\
\small{$W^{+} W^{-} Z$} & \small{$W^{+}W^{-}W^{+}$} \\
\end{tabular}
 \caption{$K-$factors, corresponding to the plots in Fig.~\ref{ptfig} }  \label{Kptfig}
 \end{center}
 \end{figure}

In Figure~\ref{ptfig}, we show results for the $p_T$ distributions of
all processes. For each phase space point, the $p_T$ of each of the
three bosons gives an entry in the histograms. The final result is
then divided by 3, yielding, as a normalization factor, the total cross section.
The corresponding K-factors are depicted in Figure~\ref{Kptfig}. 
In the $W^+ Z Z$ and $W^{+}W^{-}W^{+}$ cases, we observe
an interesting increase in the K-factor for high values of the transverse momentum.

The corresponding total cross sections are contained in Table~\ref{mix}.
As we can see the NLO corrections are quite significant, resulting to overall
K-factors of order $\sim 2$. 
 
\begin{table}
\begin{center}
\begin{tabular}{|l|c|c|c|}
\hline
Process & scale $\mu$ & Born cross section [fb] & NLO cross section [fb] \\
\hline
ZZZ & $3 M_Z$  &9.7(1)  & 15.3(1) \\
WZZ & $2 M_Z + M_W$  &20.2(1)  & 40.4(2) \\
WWZ & $M_Z + 2 M_W$  &96.8(6) & 181.7(8) \\
WWW & $3 M_W$        &82.5(5)  & 146.2(6) \\
\hline
\end{tabular}
\end{center} \caption{Cross section for the four processes, corresponding to the distributions
in Fig~\ref{ptfig}. Different values of the factorization(renormalization) scale are used for the different processes. } \label{mix}
\end{table}

\begin{figure}[ht]
 \begin{center}
   \begin{tabular}{cc}
 \begin{rotate}{90}\hspace*{35mm}{\small
 $\log(d\sigma/ d y)$} \end{rotate}
 \psfig{figure=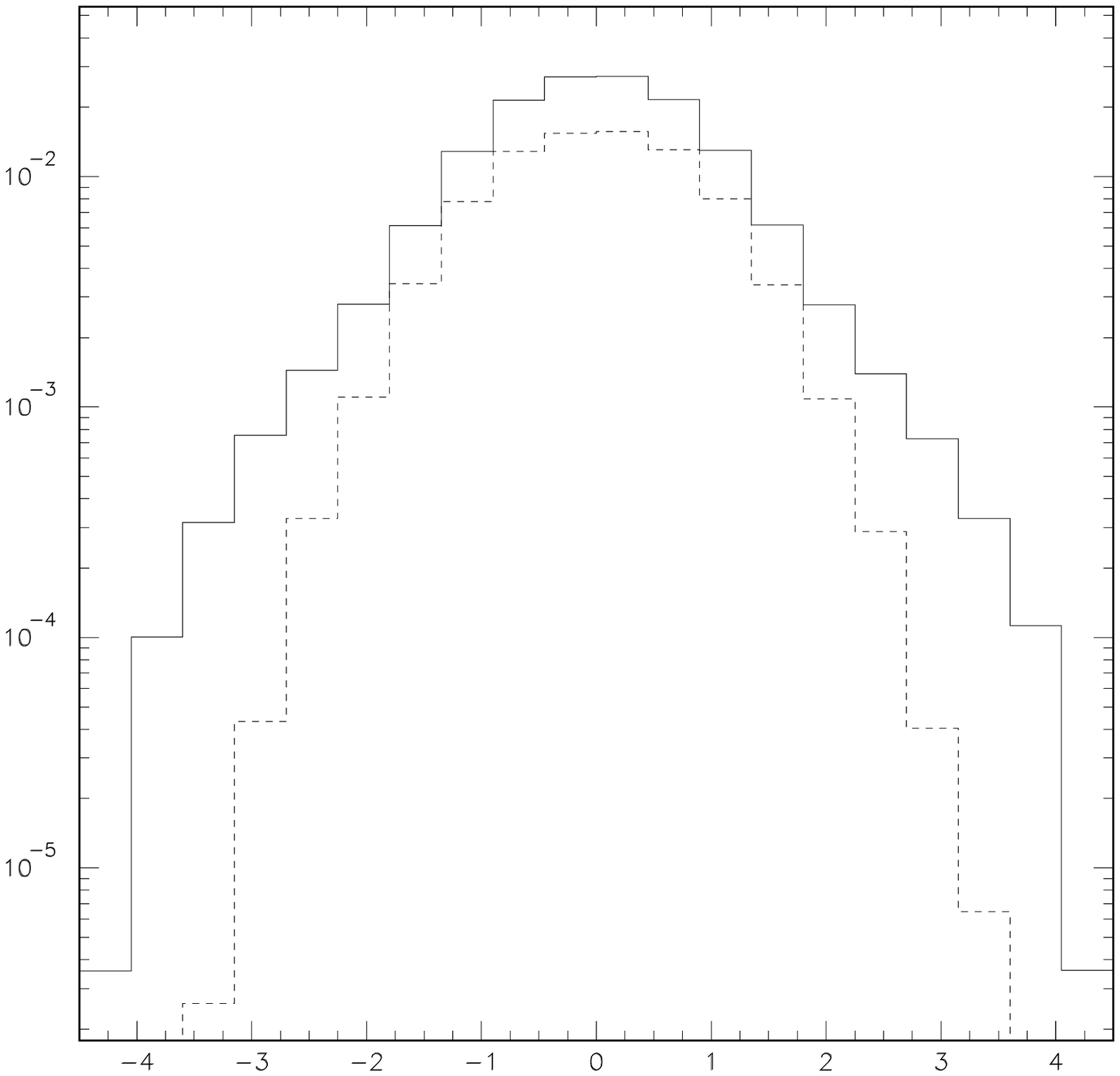,width=3.2truein} &
\begin{rotate}{90}\hspace*{35mm}{\small
 $K-$factor} \end{rotate}
 \psfig{figure=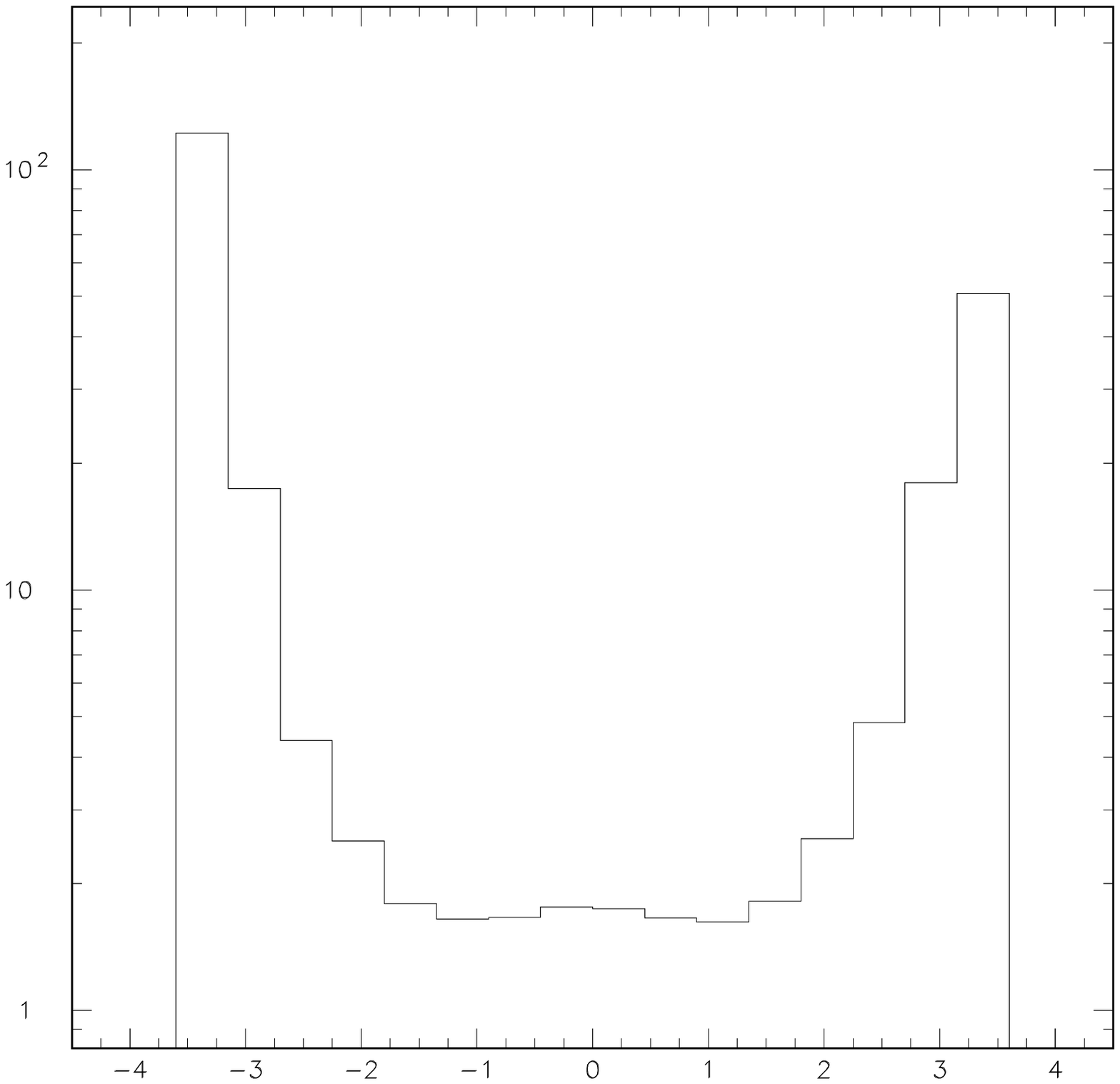,width=3.2truein}\\
 {\small $y$} &  {\small $y$}
  \end{tabular}
 \caption{Rapidity distribution, as defined in the text, for $pp \to W^{+}W^{-}W^{+}$: on the left plot, NLO (solid line)
 compared with the LO contribution (dashed line) in logarithmic scale; on the right, the corresponding $K-$factor.
 The scale is set to  $\mu=3 M_W$. }  \label{lwwwy}
 \end{center}
 \end{figure}

In Figure~\ref{lwwwy}, we show, as an indicative case, the rapidity
distribution for $WWW$ production, which is the process with the
highest cross section.
Also here, each of the three bosons gives an entry in the histograms, that
are eventually normalized to the total cross section.

 The K-factor appears to have now an
important dependence on the phase space, especially near the borders of
the available rapidity region

Let us discuss now the results obtained for the production of  $W^{+}W^{-}W^{+}$ and $W^{+} Z Z$.
In Tables~\ref{tabwww} and~\ref{tabwzz} we present the results for the cross sections (in fb)
of $pp \to W^{+}W^{-}W^{+}$  and $pp \to W^{+} Z Z$, respectively.
Each table contains the Born level, the NLO result and the corresponding K-factor.

\begin{table}[h]
\begin{center}
\begin{tabular}{|l|c|c|c|c|c|}
  \hline
  scale & $\sigma_B$ & $\sigma_{NLO}$ & K
  \\ \hline
  $\mu=M/2$ & 82.7(5) & 153.2(6) & 1.85 \\
  $\mu=M$   & 81.4(5) & 144.5(6) & 1.77 \\
  $\mu=2M$  & 81.8(5) & 139.1(6) & 1.70 \\
  \hline
\end{tabular}
\end{center} \caption{Cross section $pp \to W^{+}W^{-}W^{+}$ in fb for different values of the factorization(renormalization) scale. In the table above we set $M=3 M_Z$. } \label{tabwww}
\end{table}

\begin{table}[h]
\begin{center}
\begin{tabular}{|l|c|c|c|c|c|}
  \hline
  scale & $\sigma_B$ & $\sigma_{NLO}$ & ~K~
  \\ \hline
  $\mu=M/2$ & 20.2(1) & 43.0(2) & 2.12 \\
  $\mu=M$  &  20.0(1) & 39.7(2) & 1.99 \\
  $\mu=2M$ &  19.7(1) & 37.8(2) & 1.91 \\
  \hline
\end{tabular}
\end{center} \caption{Cross section $pp \to W^{+} Z Z$ in fb for different values of the factorization(renormalization) scale. In the table above we set $M=3 M_Z$. } \label{tabwzz}
\end{table}

\section{Summary and Conclusions} \label{conclusion}

In this paper we considered the production of three vector bosons at the LHC.
We discussed four processes, namely $Z Z Z$, $W^{+}W^{-}Z$, $W^{+} Z Z$,
and $W^{+}W^{-}W^{+}$ production: for each process we calculated the next-to-leading
order QCD corrections, presenting our results in the form of transverse
momentum and rapidity distributions.
The QCD corrections are quite sizable, with a K-factor of order 2.
The K-factor is rather uniform in $p_T$ distributions, 
while shows an important dependence on the phase space as far as rapidity 
distributions are concerned. Given the size of the corrections
the QCD corrections have to be taken into account in experimental
studies at the LHC.

This paper also represents the first complete calculation of physical cross-sections
performed using the recently introduced OPP method for the  reduction of one-loop amplitudes,
in which the reduction to known integrals is performed at the integrand level, 
using the Fortran code {\tt CutTools}.

The efficiency of the OPP method is quite good. It can be further
improved by developing the numerical evaluation of the integrand in the
one-loop amplitude by means of recursion relations~\cite{Draggiotis:2006er},
without relying on Feynman diagrams.

We conclude that the OPP method is a viable alternative to perform phenomenologically
relevant one-loop calculations, as it does not  rely on the recursive evaluation of
scalar and tensor momentum integrals. 
Its versatility and simplicity make it a very good candidate for the construction of
a universal NLO calculator/event-generator.

\bigskip

\noindent {\bf Acknowledgments} \\
 Many thanks to Andr\'e van Hameren and Achilleas Lazopoulos for numerical comparisons,
and Gudrun Heinrich for collaboration at an early stage of the project and comments
on the manuscript.
G.O. and R.P. acknowledge the financial support of the ToK Program ``ALGOTOOLS'' (MTKD-CT-2004-014319).
C.G.P.'s and R.P.'s research was partially supported by the RTN
European Programme MRTN-CT-2006-035505 (HEPTOOLS, Tools and Precision
Calculations for Physics Discoveries at Colliders).
The research of R.P. was also supported by MIUR under contract
2006020509\_004 and by the MEC project FPA2006-05294.
The research of T.B. was supported by the British Science and Technology
Facilities Council (STFC) and the Scottish Universites Physics Alliance (SUPA).
T.B., C.G.P. and R.P. thank the  Galileo Galilei Institute for Theoretical
Physics, where this work was initiated, for the hospitality and the INFN for partial support during the
completion of this work.


\end{document}